%% file: main.tex
\documentclass[sigchi]{acmart}

\usepackage[frozencache,cachedir=.]{minted}
\usepackage{balance}

\newmintinline[inlinecode]{python}{
    bgcolor=yellow!20,
    fontsize=\small
}

\AtBeginDocument{%
  \providecommand\BibTeX{{%
    \normalfont B\kern-0.5em{\scshape i\kern-0.25em b}\kern-0.8em\TeX}}}

\setcopyright{acmcopyright}
\copyrightyear{2023}
\acmYear{2023}
\acmDOI{XXXXXXX.XXXXXXX}

\acmConference[Koli Calling]{Koli Calling International Conference on Computing Education Research}{2023}{Finland}
\acmPrice{15.00}
\acmISBN{978-1-4503-XXXX-X/18/06}

\begin{document}
\title{How Novices Use LLM-Based Code Generators to Solve CS1 Coding Tasks in a Self-Paced Learning Environment}

\author{Majeed Kazemitabaar}
\orcid{0000-0001-6118-7938}
\affiliation{%
  \institution{Department of Computer Science, University of Toronto}
  \city{Toronto}
  \state{Ontario}
  \country{Canada}
}
\email{majeed@dgp.toronto.edu}

\author{Xinying Hou}
\orcid{0000-0002-1182-5839}
\affiliation{%
  \institution{School of Information, University of Michigan}
  \city{Ann Arbor}
  \state{Michigan}
  \country{USA}
}
\email{xyhou@umich.edu}

\author{Austin Henley}
\orcid{0000-0003-1069-2795}
\affiliation{%
  \institution{Microsoft Research}
  \city{Redmond}
  \state{Washington}
  \country{USA}
}
\email{austinhenley@microsoft.com}

\author{Barbara J. Ericson}
\orcid{0000-0001-6881-8341}
\affiliation{%
  \institution{School of Information, University of Michigan}
  \city{Ann Arbor}
  \state{Michigan}
  \country{USA}
}
\email{barbarer@umich.edu}

\author{David Weintrop}
\orcid{0000-0002-3009-3899}
\affiliation{%
  \institution{College of Education, University of Maryland}
  \city{College Park}
  \state{Maryland}
  \country{USA}
}
\email{weintrop@umd.edu}

\author{Tovi Grossman}
\orcid{0000-0002-0494-5373}
\affiliation{%
  \institution{Department of Computer Science, University of Toronto}
  \city{Toronto}
  \state{Ontario}
  \country{Canada}
}
\email{tovi@dgp.toronto.edu}

\renewcommand{\shortauthors}{Kazemitabaar et al.}

\begin{abstract}
As Large Language Models (LLMs) gain in popularity, it is important to understand how novice programmers use them and the effect they have on learning to code. We present the results of a thematic analysis on a data set from 33 learners, aged 10-17, as they independently learned Python by working on 45 code-authoring tasks with access to an AI Code Generator based on OpenAI Codex. We explore several important questions related to how learners used LLM-based AI code generators, and provide an analysis of the properties of the written prompts and the resulting AI generated code. Specifically, we explore \textbf{(A)} the context in which learners use Codex, \textbf{(B)} what learners are asking from Codex in terms of syntax and logic, \textbf{(C)} properties of prompt written by learners in terms of relation to task description, language, clarity, and prompt crafting patterns, \textbf{(D)} properties of the AI-generated code in terms of correctness, complexity, accuracy, and \textbf{(E)} how learners utilize AI-generated code in terms of placement, verification, and manual modifications. Furthermore, our analysis reveals four distinct coding approaches when writing code with an AI code generator: \textit{AI Single Prompt}, where learners prompted Codex once to generate the entire solution to a task; \textit{AI Step-by-Step}, where learners divided the problem into parts and used Codex to generate each part; \textit{Hybrid}, where learners wrote some of the code themselves and used Codex to generate others; and \textit{Manual} coding, where learners wrote the code themselves. The \textit{AI Single Prompt} approach resulted in the highest correctness scores on code-authoring tasks, but the lowest correctness scores on subsequent code-modification tasks during training. Our results provide initial insight into how novice learners use AI code generators and the challenges and opportunities associated with integrating them into self-paced learning environments. We conclude with various signs of over-reliance and self-regulation, as well as opportunities for curriculum and tool development.
\end{abstract}



\maketitle

\input{sections/1-introduction}
\input{sections/2-related-work}
\input{sections/3-methodology}
\input{sections/4-results}

\input{sections/5-discussion}
\input{sections/6-limitations}
\input{sections/7-conclusion}

\bibliographystyle{ACM-Reference-Format}
\bibliography{references}

\end{document}

%% file: sections/1-introduction.tex
\section{Introduction}

Large Language Models (LLMs) trained on code like OpenAI Codex \cite{chen2021evaluating} are capable of generating functioning programs from natural language descriptions. Since publicly made available by companies like OpenAI through user-facing tools such as ChatGPT (a Q\&A chatbot) or Github Copilot (an IDE-based AI coding assistant), the code generation capabilities of LLMs are becoming more accessible to a wider array of people. These tools have the potential of scaling up computing education in self-paced learning environments and broadening participation in computing by assisting beginners with debugging, code generation, code explanation, and responding to questions about code. 

Despite their potential benefits, LLMs present challenges in educational contexts. Their usage could result in learner dependency, hindering code authorship without assistance. Novice coders may also struggle with technical jargon, expressing coding intent, and comprehending or verifying AI-generated code. Additionally, from an educator's perspective, issues around academic integrity and plagiarism pose valid concerns \cite{becker2023programming}.


While LLM-based code generators are not specifically designed for education contexts, recent studies have assessed their performance in computer science education and their effects on learners and teachers \cite{denny2023conversing, finnie2022robots, finnie2023my}. Kazemitabaar et al. \cite{kazemitabaar2023studying} compared students' learning outcomes with and without an AI code generator based on OpenAI Codex. They found that using Codex did not harm learners' performance on post-tests one week later, and students who performed better on Scratch pre-tests performed significantly better on retention post-tests if they had prior access to Codex. However, we still do not know how these outcomes occur. To fully understand the benefits of AI code generators in education, it's crucial to know how novices use them while learning to code \cite{becker2023programming, finnie2023my}.

In this paper, we conduct a thematic analysis on a data set of 33 novice learners (ages 10-17) who had access to an AI Code Generator while learning Python programming for the first time. We attempt to answer two main research questions:
\begin{itemize}
    \item \textbf{RQ1:} How do novices use and interact with LLM-based Code Generators when learning to write code by practicing CS1 coding tasks in a self-paced learning environment? \textbf{A:} When do learners use Codex in the problem-solving process? \textbf{B}: What are learners asking Codex to generate? \textbf{C}: What are the properties of the prompts that novices craft in terms of relationship to the task description, language, and vagueness? \textbf{D}: What properties does the AI-generated code have? \textbf{E}: How do novices use, modify, and verify AI-generated code?
    \item \textbf{RQ2:} What coding approaches do novice learners employ when they have access to LLM-based code generators to solve programming problems? How do these approaches differently impact learning outcomes measured by retention post-tests?
\end{itemize}

To answer these questions, we analyzed log data from a previous study \cite{kazemitabaar2023studying} using a custom log analysis tool. Our findings reveal when and how novice learners interact with LLM-based code generators in addition to four distinct coding approaches that reflect learners’ personal choices for utilizing AI code generators during programming. Our thematic analysis reveal various signs of over-reliance and self-regulation in students when they write code with an AI code generator. Our results suggest the importance of effective usage patterns for maximizing learning outcomes when novice learners have access to AI code generators. These findings can inform the design of future introductory programming tools, and highly scalable self-paced learning environments that incorporate AI code generators and the pedagogy that accompanies them.

%% file: sections/2-related-work.tex
\section{Related Work}

\subsection{Programmers' Experience with AI Coding Assistants}

Large language models (LLMs) are deep neural networks trained on extremely large-scale model parameters using terabytes of textual data sets. When trained on large corpora of source code, these models can also generate code from natural language descriptions \cite{austin2021program, chen2021evaluating, jain2022jigsaw} . Pre-trained LLMs, like OpenAI Codex \cite{chen2021evaluating} based on GPT-3, Google PaLM \cite{chowdhery2022palm}, and DeepMind AlphaCode \cite{li2022competition}, have enabled LLM code generation tools like Github Copilot \cite{copilot}. The emergence of these technologies has sparked research on evaluating how professional programmers utilize AI code generators \cite{sarkar2022like}. 

One line of research is exploring how users use GitHub Copilot \cite{barke2023grounded, mozannar2022reading, tang2023empirical, vaithilingam2022expectation}, a commercialized programming tool powered by OpenAI Codex. GitHub Copilot can provide autocomplete-style code suggestions based on existing code and natural language descriptions (e.g., comments). For instance, Barke et al. \cite{barke2023grounded} conducted a grounded theory analysis with 20 programmers to identify two ways in which programmers interact with Copilot: acceleration, where Copilot speeds up the writing of code in "small logical units" because the programmer has a clear intention, and exploration, where Copilot suggestions are used to help with planning and exploration. There also have been other studies analyzing programmer experiences with more experimental AI coding assistants \cite{jiang2022discovering, xu2022ide, xu2019block}. For example, Ross et al. \cite{ross2023programmer} developed a conversational-based coding assistant and found two main usage patterns: bringing assistant’s help to solve the entire challenge at once, or breaking it down to solve each smaller task individually. 

However, most of the existing usability work on AI-assisted coding assistants focuses on the experience of expert programmers within well-designed experimental context, typically by testing a small number of programming tasks. To our knowledge, only Jayagopal et al. \cite{jayagopal2022exploring} focused novice programmers’ experience with different program synthesizers including Copilot, but only for three tasks.  To further elucidate this topic, we believe it would be enlightening to understand novice programmers’ interaction with such AI code generators in an authentic programming context with a larger number of coding tasks. Therefore, in this work, we would like to analyze the user interactions collected in a data set of novice learners as they make progress and learn Python programming in a self-paced learning environment while they have access to an AI code generator.

\subsection{Large Language Models in Computer Science Education}

As LLMs become more widely used in practice, education researchers are exploring the potential of LLMs to produce educational content, enhance student engagement and customize learning experiences \cite{kasneci2023chatgpt}. Recent work in the computer science education community has started to explore the implications and opportunities of LLMs on computer science learning from different perspectives \cite{becker2023programming}. Most of the recent work focused on understanding the capabilities of LLMs for completing programming tasks \cite{denny2023conversing}, generating instructional content \cite{leinonen2023using}, and developing new content creation methods \cite{denny2022robosourcing}. For example, Finnie-Ansley et al. \cite{finnie2023my} showed that Open AI Codex performs better than most students on code writing questions in both CS1 and CS2 exams. When it comes to creating CS educational content, prior evaluations showed that LLMs have the potential be used to produce high enhancements on programming error messages \cite{leinonen2023using} and provided high-precision feedback on code for fixing syntax errors \cite{phung2023generating}. Furthermore, Sarsa et al. \cite{sarsa2022automatic} analyzed the novelty, plausibility, and readiness of 120 programming exercises generated by OpenAI Codex and proposed the potential of using such models to come up with coding assignments. As for code explanations, MacNeil et al. \cite{li2022competition} reported student experience with LLM-generated code explanations in a web software development e-book. They showed that most students perceived the code explanations as helpful, but engagement depends on code length, code complexity and explanation types.  

To better integrate LLMs into computer science education and achieve student-centered learning, it is important to understand how novice students use and interact with AI code generators while they learn to code. For instance, we need to know what usage patterns and coding approaches students employ when learning with AI code generators, how these patterns and approaches impact their learning, what types of prompts students write, what are the attributes of the generated code and its relationship to the prompts, and how students integrate AI-generated code into their existing code. In this paper, we investigate these questions through a systematic analysis of student log data on their interaction with OpenAI Codex.

\subsection{Supporting Novices While Writing Code}

To help students write code, previous research has explored a variety of scaffolded approaches. Bruner \cite{bruner1966toward} coined the term "scaffolding" to refer to the process of giving students support structures so they can learn a subject or skill that is above their level. Appropriate scaffolding will equip learners with sufficient knowledge and skills to perform the task independently. One direction provides support before writing the actual code. For example, flowcharts have been used to brainstorm and organize solution ideas before diving into coding, according to Renske \& Sjaak \cite{smetsers2017problem}. This has resulted in an improvement of algorithm design and programming skills. Moreover, Cunningham et al., \cite{cunningham2021avoiding} described a multi-stage programming process including arranging plans explicitly.

Another direction is to provide immediate assistance during the actual code writing process. Such immediate assistance can be done by providing detailed feedback based on the student's current code states \cite{keuning2018systematic}. This feedback could be explanations on what is wrong in the existing code \cite{singh2013automated}, suggestions on how to fix the error \cite{sykes2010design}, or next-step hints on improving the student's solution towards the goal \cite{rivers2017data}. In addition, when students ask for help, the available support could also be a library of worked examples that students can run and modify with output \cite{wang2022exploring} or an equivalent Parsons problem that students can actively work on \cite{hou2022using}. While previous research has examined a wide range of supportive methods, none of them have considered the implications of AI code generator to support students to write code.

As a first step, Kazemitabaar et al. conducted a controlled experimental to investigate the effects of incorporating an AI code generator during code writing compared to a baseline condition (writing code without any assistance) \cite{kazemitabaar2023studying}. The previous work highlighted the need for a systematic analysis of students' interaction with AI code generators to better understand their impact on learning. Building upon this prior research \cite{kazemitabaar2023studying}, our aim is to identify key moments when students choose to use AI code generators, the attributes of the code they seek in different situations, the language properties used in prompts, how they utilize AI-generated code, and how the coding approaches they use with AI code generators, influence their learning.

%% file: sections/3-methodology.tex
\section{Methodology}

\subsection{Data Set and Data Instrumentation}

To explore how novice use AI code generators, we analyzed a data set from a prior study \cite{kazemitabaar2023studying} in which 69 novice learners (ages 10-17) used Coding Steps (https://github.com/MajeedKazemi/coding-steps) as part of a three-week study. Coding Steps is a self-paced, online learning environment that includes 45 CS1 Python coding tasks that were designed to gradually introduce new concepts. The system includes an embedded programming environment, functionality to submit code to remote instructors that grade submitted work and provide real-time personalized feedback, a novice-friendly Python tutorial, and an AI code generator (based on OpenAI Codex).

The original study included two conditions: one in which learners had access to an AI code generator (the Codex condition), and another in which the AI code generator was not available (the Baseline condition). In this work, we reuse the same data but only focus on the log data from the 33 learners in the Codex condition to investigate new research questions. Four types of timestamped logs were collected from learners: code edit logs, console run logs, AI code generation logs (including prompt message and generated code), and submission logs. Log data has become an important data source to understand programming experiences \cite{brandt2009two} and coding approaches \cite{finnie2023my, ichinco2015exploring}. Log data can provide valuable insights for computing education researchers \cite{marwan2020unproductive} as it can capture detailed information about students' experiences and actions.

\subsection{Original Study Design and Participants}
The original study consisted of ten 90-minute sessions, held over three consecutive weeks, covering a session of introduction to Scratch, seven sessions of Python training, and two evaluation sessions including a retention post-test one week later.

\subsubsection{Participants}
The original study included 69 participants (21 female, 48 male) aged 10-17 (M=12.5; SD=1.8). Participants were recruited from over 200 sign-ups at coding camps in two major North American cities. The study was approved by the Research Ethics Board of the original author's institution. Parental/guardian consent was obtained before the first session and each participant received a \$50 gift card as compensation.

From the 33 participants that we focus on in this study, none reported any prior text-based programming experience, 32 indicated using a block-based programming environment like Scratch or Code.org, and 12 indicated taking a programming-related class in the past. In terms of language, 25 participants were English speakers, while 1 reported difficulty explaining things in English. More details about the study can be found in the paper that presents the original study \cite{kazemitabaar2023studying}.

\subsubsection{Study Procedure}

The first session included a one-hour lecture on the fundamentals of programming using Scratch, and a pre-test evaluation including 25 multiple-choice questions about Scratch programming. The Scratch lecture covered all the key concepts required in the training phase, including: input/outputs, random numbers, arithmetic operators, conditionals, comparators, logical operators, loops, and arrays.

Students then began the training phase by watching an introduction video with several examples of AI code generation. As a self-paced online learning environment, learners received personalized feedback from remote instructors and were provided a novice-friendly Python documentation that included worked examples, common Python errors, and debugging strategies. During this phase, learners worked on 45 two-part programming learning tasks and 40 multiple-choice questions at their own pace using the Coding Steps IDE (Figure 1). The tasks and the topics were presented in a fixed order with gradually increasing complexity. For more details about each of the tasks, read the Appendix A of the original paper \cite{kazemitabaar2023studying}.

Each coding task had two parts: code-authoring and code-modification. For each part, learners first read the task description and a few examples of input and expected output. To finish a code-authoring task, learners were instructed to write code in the code editor with access to the AI code generator. Learners were also provided with a Python documentation inside the IDE. Regarding the code-modification tasks, learners were given their own accepted submission from the code task as the starting point. If the learner skipped a code-authoring task, they were shown an example correct solution as the starting point for the associated code-modification task. Learners had no access to the AI code generator for modification tasks.

Finally, the evaluation phase comprised an immediate post-test followed by a retention post-test administered one week later. The tasks used in the evaluation phase were analogous in terms of difficulty and topics to the tasks used in the learning phase. Both post-tests contained 10 coding tasks (5 code-authoring and 5 code-modification tasks), and 40 multiple-choice questions. Learners had no access to the AI code generator, Python documentation, or instructor feedback during the evaluation phase.

\begin{table*}[t]
\centering
\caption{High-level codebook that was used to analyze each Codex usage.}
\label{tab:highlevel_codebook}
\begin{tabular}{p{0.37\linewidth} p{0.59\linewidth}}
    \toprule
    \textbf{High-Level Code Dimensions} & \textbf{Sub-dimensions} \\
    \midrule
    Context of Codex Usage (RQ1 \textbf{A}) & \textit{prior manual edits}, \textit{prior codex usage}, \textit{existing issues}\\
    Prompt Attributes (RQ1 \textbf{B}, \textbf{C}) & \textit{request content}, \textit{relationship to task}, \textit{vagueness}, \textit{requesting syntax/logic}, \textit{fixing/localizing issue}, \textit{repeated prompt} \\
AI-Generated Code Properties (RQ1 \textbf{D}) & \textit{quality properties}, \textit{quality reason}, \textit{generated extra unspecified code}, \textit{outside curriculum} \\
Using and Modifying AI-Generated Code (RQ1 \textbf{E}) & \textit{placement}, \textit{modifying existing code}, \textit{modifying AI-generated code}, \textit{testing/verification}, \textit{next action} \\
  \bottomrule
\end{tabular}
\end{table*}

\subsection{Data Analysis}

To assist with our data analysis, we developed a web interface for visualizing log data and replicating student behaviors in a vertical time sequence based on log entries. The interface displays high-level metadata for each task, keystroke counts for edit actions, side-by-side diffs for code modifications, prompt messages along with AI-generated code for Codex usages, and console input/output for execution actions. When analyzing data, we applied a combination of deductive and inductive approaches in thematic analysis \cite{bernard2016analyzing, miles1994qualitative}. We applied a deductive approach to categorize learners' log data into groups based on task number and student ID, then we analyzed each student-task pair under two levels.

For analysis, we defined each AI code generation request (which was initiated by a prompt) as a single unit of analysis, resulting in a total of 1666 data points. To answer RQ1, we created four high-level code dimensions: (i) the context of using Codex in terms of editor contents and prior actions (RQ1 \textbf{A}), (ii) prompt attributes including requested content, details, and language (RQ1 \textbf{B}, \textbf{C}), (iii) quality of the AI-generated code (RQ1 \textbf{D}), and (iv) utilizing AI-generated code (RQ1 \textbf{E}). This enabled us to focus on relevant data for each research question in following rounds of analysis \cite{bingham2021deductive}. Additionally, to answer RQ2, we derived four high-level recurring coding approaches based on RQ1 and then assigned a single approach to each of the tasks submitted by learners. 

Under each high-level dimension, we applied an inductive approach where we read through all the data and allowed codes to emerge during the process \cite{bingham2021deductive}. We generated the codebook iteratively following this process: First, two researchers (the first two authors of this paper) familiarized themselves by going through all the logs, identifying candidate sub-dimensions (if possible), and specifying codes for each dimension. These codes were then iteratively revised as the data was reviewed and drafted into an initial codebook. Moreover, the two researchers independently coded data for three tasks (6.4\% of the data), each with different difficulty levels and topics using the initial codebook. After that, they discussed the initial coding results, resolved conflicts, and further refined the codebook.

After refining the codebook (Table \ref{tab:highlevel_codebook}), the two researchers independently coded another five tasks (13.2\% of the data) and reached an inter-rater reliability of 0.87 (alpha values > 0.80 \cite{neuendorf2017content}) using percentage agreement \cite{miles1994qualitative}. After addressing the disagreements and finalizing the codebook, the two researchers coded the remaining data individually by assigning task numbers at random. The full codebook can be found at https://tinyurl.com/codex-analysis-codebook.

%% file: sections/4-results.tex
\section{Results}
Overall, we analyzed 1666 Codex usages from 1379 submitted tasks (356 tasks were submitted without using Codex). A Codex usage is designated by writing a prompt in the AI code generator's text-box and pressing the generate button. The generated code is automatically placed at the user’s cursor in the editor. We define the problem-solving process as a sequence of actions until the final submission of a task. Actions include manually editing code, using Codex, using the documentation, running the code, providing input to it, and submitting it for evaluation by remote instructors.

We denote participant numbers as $P_n$ (ranked based on pre-test Scratch scores: $P_1$ with the highest score).

\subsection{RQ1 A: When do Learners Use Codex?}
We identified five primary scenarios (based on the state of the editor) in which learners prompted Codex: at the beginning (46\%, n=760), after clearing the editor (5\%, n=83), after manual coding (17\%, n=282), after using Codex (34\%, n=572), and while already having the solution (1\%, n=16). Below we briefly describe each scenario.

\subsubsection{Starting with Codex} 
In 760 tasks, learners used Codex at the beginning of the task, often (92\%, n=703) with no initial attempt of prior manual coding. Common behaviors included copying the full task description to generate the entire solution (66\%, n=503), rephrasing task description to generate the entire solution (7\%, n=53), and breaking down the task into subgoals (26\%, n=199).

\subsubsection{Using Codex After Clearing the Editor}
In 83 Codex usages, learners cleared their editor after being stuck with invalid code with repeated compiler errors and wrong edits (86\%, n=71), or failing to properly test correct code (14\%, n=12). Learners then cleared their editor and prompted Codex for either the entire (51\%, n=42) or part of the task (49\%, n=41). Notably, 75 of these 83 instances (90\%) occurred after unsuccessful Codex usages, resulting in broken AI-generated code or misplaced code that was challenging to fix manually.

\subsubsection{Using Codex After Manual Coding}
Out of 293 Codex usages, learners used Codex following instances of manual coding. In 102 (35\%) instances, learners manually wrote correct code prior to prompting Codex. Learners typically wrote declarations (e.g., getting input from user or generating random numbers) and used Codex to implement the next major subgoal. Six instances involved learners writing complex code with loops or nested conditionals prior to prompting Codex. Furthermore, prior to 115 (39\%) Codex usages, learners authored mostly correct code with minor issues. Common patterns included (i) struggling with string and integer concatenation before prompting Codex to fix, (ii) 28 instances of deliberately writing incomplete code like \inlinecode{num =} before prompting Codex with “random number”. In two unique cases, $P_3$ manually wrote code that included ellipsis like \inlinecode{print("It took ", ..., "attempts.")} and then prompted Codex to fill in the behavior (e.g., “counting attempts”). In 76 (26\%) instances, Codex was used after writing mostly incorrect code. Examples included syntax errors in code involving string and variable concatenation such as \inlinecode{join(name + "bot")} ($P_2$), obtaining number from the user like \inlinecode{print "Enter a number:"} ($P_{25}$). Learners used Codex to fix these issues.

Furthermore, in the 191 cases where learners faced issues after manual coding, Codex was used to fix existing code in 44\% (n=85) instances, generate the entire solution in 18\% (n=34) cases, or generate new code ignoring the issue in 32\% (n=62) cases. These initial manual coding attempts before using Codex highlight self-regulation in learners. Moving forward, it is crucial for future tools and curriculum to prioritize effective methods for using LLMs for debugging manually written code.

\subsubsection{Using Codex after Using Codex}
In 572 instances of using Codex after a previous Codex usage, learners handled the prior AI-generated code as follows: 53\% (n=304) kept it unchanged, 36\% (n=206) deleted it, 7\% (n=41) made minor modifications, and 4\% (n=21) broke it. Among the cases where learners correctly placed the AI-generated code (304 cases), 80\% (n=243) proceeded to generate new subgoals. Of interest, we identified 84 Codex usages in which learners requested code similar to what was already in their editor. Furthermore, after deleting the prior AI-generated code (206 cases), learners either re-attempted using Codex by rephrasing the prompt message in 40\% (n=130) cases, or decided to clear their entire editor (32\%, n=66).

\subsubsection{Using Codex while Having Solution}
We identified 16 instances where learners used Codex when they already had the task's solution. Of interest, in six cases, learners compared their manually written solution with an AI-generated solution and performed minor edits to their own solution.

\begin{figure*}
    \centering
    \includegraphics[width=\textwidth]{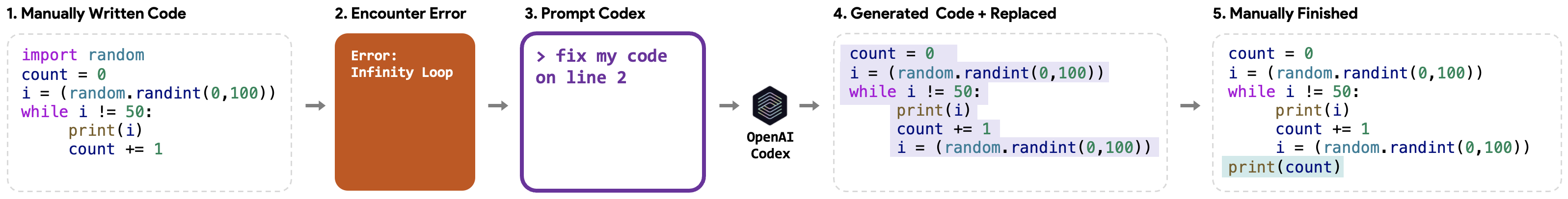}
    \caption{An example of using AI-generated code as an example to fix syntax error with writing loops.}
    \Description{An example of using AI-generated code as an example to fix syntax error with writing loops.}
    \label{fig:explicit_fix}
\end{figure*}


\subsection{RQ1 B: What are Learners Asking from Codex?}
Learners used Codex to generate the entire solution (43\%, n=723), generate new subgoals (37\%, n=626), and fix existing code (7\%, n=110). When generating new subgoals, we found 135 Codex usages where learners used Codex to generate code similar to existing code in their editor which may indicate over-reliance on the AI code generator. For example, when $P_{25}$ prompted Codex with “ask for another number” they already had similar code in their editor: \inlinecode{number = int(input("Enter a number: "))}. Additionally, when fixing code using Codex, learners were often (61\%, n=73) able to correctly localize their issues. In seven cases, learners explicitly prompted Codex to fix their code (see Figure \ref{fig:explicit_fix} for an example).

By analyzing crafted prompts, we identified two categories: requesting either \textbf{pure syntax} (e.g., \textit{"get a number from user"}), or both \textbf{syntax and logic} (e.g., \textit{"get a number until it's 0"}). Of the 723 prompts where learners prompted Codex to generate the entire solution, 46\% requested \textit{syntax and logic}, while 54\% sought \textit{pure syntax}. However, in the 626 prompts for new subgoals, 85\% asked for \textit{pure syntax}, with the rest requesting \textit{syntax and logic}. This demonstrates that learners mostly requested syntax when they decomposed a task into multiple subgoals.

\subsection{RQ1 C: Prompt Properties}
For each Codex usage, we analyzed the properties of the prompt in terms of its relationship to the task description, its language, and its clarity. Additionally, our analysis highlights two prompt crafting patterns: (i) solving a task by dividing its task description sentence by sentence for each prompt, and (ii) using repeated and slightly reworded prompts to achieve desired code.

\subsubsection{Prompt Relationship to Task Descriptions}
More than half (52\%, n=864) of all requested prompts were copied directly from the task description, with 27\% (n=233) being partial copies (requesting code for part of a task). The major groups of the remaining prompts (48\%, n=802) were written independently by learners: (i) 335 prompts were accurately reworded versions of the task description with all required details and using similar vocabulary, (ii) 223 prompts included less details or vague interpretations of the task which often led to code with omitted details or unintended behavior, (iii) 41 were based on incorrect task interpretations, and (iv) 71 unrelated to the task or Python programming.

\subsubsection{Prompt Language}
We identified 86 prompts written by learners that were similar to a pseudocode and included syntactical elements. For example, when $P_4$ wanted to find the largest number in a list, they prompt Codex with a pseudocode that specified the exact logic \textit{“for n in numbers, if n > l, set l to n”} followed by \textit{“print Largest number: l”}. Compared to other learners who simply prompted the behavior (e.g., \textit{“find the largest number”}), such usages of pseudocode indicate a deeper algorithmic thinking. Additionally, learners sometimes articulated their desired code by specifying the output format such as \textit{“display message Grade: grade”}. We also observed prompts that described the entire behavior through an input/output example, including when $P_{12}$ prompted \textit{“output: What is your name? output: Hello, Bob!”} which generated the correct code. Furthermore, sometimes learners skipped static values in their prompt messages so they could focus more on the logic while crafting prompts. For example, $P_6$ prompted \textit{“if variable number is greater than 75 print 2. If anything else print 3”} and then modified the values of “2” and “3” in the AI-generated code based on the task requirements.

\subsubsection{Prompt Clarity}
We found that about 28\% (n=201) of prompts had some indicators of vagueness such as being under-specified. A common theme was not specifying the initial value for variable declarations like \textit{“variable called name”} ($P_8$, $P_{10}$, $P_{16}$). To generate random numbers, vague prompts did not indicate where to store the number, or the range to select from like \textit{“rand”} ($P_3$), or \textit{“number = roll”} ($P_{32}$). Vague prompts for conditionals did not specify the condition like \textit{“check variable”} ($P_{25}$). To define loops, vague prompts did not specify the stop condition like \textit{“forever, set M to input”} ($P_4$), or were expressed vaguely like when $P_{20}$ prompted \textit{“go back to top”} after noticing their code does not repeat.

\subsubsection{Prompt Crafting Patterns}
We found that learners frequently decomposed tasks into individual prompts. While effective for sequential and independent subgoals, this approach faltered when a sentence in the task altered or built on top of the instructions given in the previous sentence. In the latter case, the LLM usually regenerates the existing code with the added functionality, therefore, requiring learners to replace it with their existing code. Moreover, we examined tasks with multiple prompts and observed that 7\% (n=109) were exact repetitions, and 3\% (n=55) were slight rephrasings, typically used when initial prompts failed to yield the desired code. Only in 13 Codex usages learners added meaningful detail in reworded prompts. Notably, we found six instances where learners updated minor values via reworded prompts rather than modifying values in the pre-existing AI-generated code.

\subsection{RQ1 D: AI-Generated Code Properties}
We analyzed AI-generated code from three dimensions: correctness, relation to curriculum (complexity), and relation to prompt (accuracy). We found that of the 1666 Codex usages, 81\% (n=1357) of all AI-generated code was produced without any identifiable problems, while 19\% (n=309) of the generated code exhibited some problematic characteristics, including: not following task requirements 28\% (n=87), regenerating existing code 28\% (n=86), missing minor code 19\% (n=60), incorrect code 8\% (n=25), and generating only comments 8\% (n=25). Finally, we report on the relationship between the quality of AI-generated code and the quality of the prompts associated with them.

\subsubsection{Correctness of AI-Generated Code}
Here we focus on two problematic characteristics of AI-generated code. First, Codex regenerated code that existed in the editor in 86 usages. This behavior caused complications when learners had an erroneous code and used Codex to fix it. For example, when $P_{15}$ wanted to fix their incorrect code \inlinecode{message = ("num is:" + num)}, they prompted Codex with \textit{“make a variable called message and define it so that it means num is: the num variable”}, however, Codex regenerated their incorrect code. Furthermore, in 60 cases Codex did not generate minor pieces of code that were necessary for the code to execute properly, which was often (56 cases) a missing import random statement, or not casting input from string to integer.

\subsubsection{Complexity of AI-Generated Code}
We identified 22 instances in which Codex generated code that was either not covered in the curriculum or from advanced topics that were supposed to be introduced later in the training phase. For example, when learners have not reached tasks on loops, $P_{26}$ prompted Codex with \textit{“user must enter a value”} which resulted in Codex generating a while loop that repeatedly received an input from the user until it had a valid value. This is probably because the student included the word \textit{“must”} in their prompt. Of interest, 11 usages in which the learner copied the task from the task description yielded into Codex generating code that was outside of the curriculum.

\subsubsection{Accuracy of AI-Generated Code}
We identified 204 cases in which Codex produced additional code that was unspecified in the prompt message, and instead was predicted from the editor’s content or the provided prompt message. For example, when the editor contained three variables called \inlinecode{num1}, \inlinecode{num2}, and operator, $P_3$ prompted Codex with \textit{“check operator symbol”} which generated four if-else conditions checking the operator variable with the basic math operators.

In 39 cases, the predictions of the extra code were first generated as comments followed by some code. For instance, when $P_{25}$ prompted Codex with \textit{“check if a number is divisible”}, it first generated \inlinecode{\# by 3} and then on the next line, \inlinecode{if number \% 3 == 0:}.

We analyzed the extra code that was generated in each of the 204 cases and compared it to the requirements of its corresponding task to determine the usefulness of the extra code. We found that 34\% (n=70) were wrong and had to be deleted, 34\% (n=60) were directly usable without no modification, 14\% (n=29) were usable after some manual modifications, and 12\% (n=25) were usable after minor changes to the values.

\subsubsection{Effect of Prompt Quality}
We analyzed the context, prompt, and the generated code to understand why Codex generated low-quality code. We identified reasons including poorly crafted prompts that do not properly articulate the intended behavior (n=105), prompts with missing important detail (n=71), and existing low-quality code in the editor causing Codex regenerate similar code (n=34).

Furthermore, by analyzing the source of prompts, out of 864 prompts that were copied from the task description, 81\% (n=701) produced high-quality code, 7\% (n=59) prompts produced extra unspecified code, 5\% (n=41) produced code with a minor missing item, while 7\% (n=63) were unusable as it either repeated existing low-quality code, generated only comments, or the code was not following the task requirements. From 354 prompts that were reworded versions of the task description, 72\% (n=256) generated high-quality code, 10\% (n=38) generated extra unspecified code, 2\% (n=9) generated code with a minor missing item, while 15\% (n=51) were incorrect, only comments, or were repeated incorrect codes. From the 225 prompts that were reworded versions of the task description but with less detail, 46\% (n=104) generated high-quality code, 20\% (n=66) generated extra unspecified code, 2\% (n=4) with minor missing items, while 22\% (n=50) prompts generated low-quality code that were not usable. From the 89 prompts that included pseudocode and syntactical elements, 63\% (n=56) produced high-quality code, 16\% (n=14) generated extra unspecified code, 20\% (n=18) generated code that was not the intended behavior.

\subsection{RQ1 E: Utilizing AI-Generated Code}

Here we report how learners utilized AI-generated code in terms of \textit{placement}, \textit{verification}, and \textit{modification}. We also present a pattern in which learners use AI-generated code as an example to fix their incorrect code.

\subsubsection{Placement of AI-Generated Code}

The AI code generator in Coding Steps generates code at the user's cursor location. Therefore, learners had to manually adjust the code placement if their cursor was in the wrong spot. Additionally, when the AI-generated code included parts of the existing code, learners also needed to replace their existing code with the AI-generated code. 

For the most part, learners were able to place the AI-generated code in the correct position. However, out of 1666 Codex usages, we identified 69 instances in which learners placed the AI-generated code incorrectly. There were seven cases of not placing the AI-generated code at the correct indentation (e.g., inside a loop), and six instances of placing the AI-generated code before a variable that it accessed was declared. Additionally, in 18 cases, instead of replacing their original code with the newly AI-generated code, learners kept both versions.

\subsubsection{Verifying AI-Generated Code}

Through our thematic analysis, we identified three active verification approaches including tinkering with AI-generated code, properly running the code to evaluate it, and manually adding code for verification.

We identified 30 instances in which learners actively tinkered with the AI-generated code by temporarily changing values, conditions of while loops, modifying syntax, or even temporarily removing code to see how it contributed to the entire program. For example, after $P_3$ declared a list of five elements, they prompted Codex with \textit{“print 1st message in list”} and Codex generated \inlinecode{print(myList[0])}. They then tinkered with the list index, temporarily changing it from 0 to 1, testing the code to see how it affects the output.

From the total 1666 Codex usages, learners tested their AI-generated code in 60\% (n=1005) usages. Of which, 71\% (n=720) were executed correctly without any errors, and 29\% (n=285) with errors or incorrect behaviors. However, after 485 Codex usages, learners did not run their AI-generated code to test it. In 65 cases (13\%) learners deleted the AI-generated code, in 166 cases (34\%) learners moved on to using Codex for the next part of the task, and 63 cases (13\%) in which learners submitted the AI-generated code as their final solution without running the code.

Future studies with more rigorous techniques such as think-aloud or eye-tracking studies are required to properly understand how learners verify and check AI-generated code.

\begin{figure*}
    \centering
    \includegraphics[width=\textwidth]{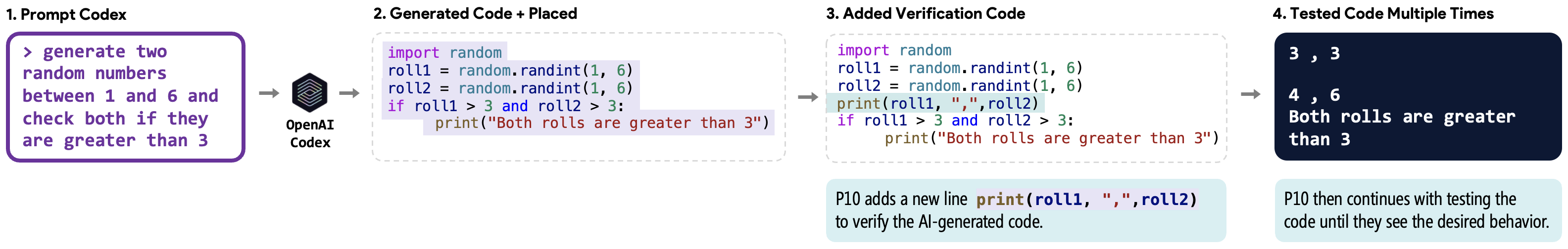}
    \caption{An example of keeping the original code instead of replacing it with AI-generated code ($P_{12}$).}
    \Description{An example of keeping the original code instead of replacing it with AI-generated code ($P_{12}$).}
    \label{fig:adding_verification}
\end{figure*}

Furthermore, we observed that some learners added code manually to verify the behavior of the AI-generated code on several tasks. For instance, out of the 22 learners who used Codex for a task that required counting the digits of a randomly generated number, the AI-generated code did not include a print statement to display the random number, only displaying the final digits count. Therefore, it made sense for learners to verify the code before submitting it. Among the 11 learners who used Codex by copying the task description, only $P_{14}$ and $P_{16}$ correctly added verification code, while four faced difficulties. Furthermore, out of the seven learners who used Codex after manual coding, five correctly added verification code, one struggled, and one did not verify the code at all. Figure \ref{fig:adding_verification} displays an example from $P_{10}$  manually adding verification.

\subsubsection{Modifying AI-Generated Code}

We identified 175 cases in which learners correctly modified the AI-generated code to align it better with the task’s requested behavior. Some edits were towards getting the correct output format. For example, on a task in which learners used Codex which did not produce the correct output format, 10 learners had to modify \inlinecode{message = num1 * num2} into \inlinecode{message = "num1 times num2 is " + str(num1 * num2)}. However, in 57 instances, learners broke correct AI-generated code and were not able to fix it. In some cases, such incorrect edits were accidents (e.g., removing a parenthesis) which left them confused for the rest of the task. In other cases, learners failed to get the desired output format and instead broke the AI-generated code. For example, when $P_{15}$ tried to add \inlinecode{"Length:"} to the beginning of \inlinecode{print(len(my\_list))} , they instead wrote: \inlinecode{print(len("Length:" + my\_list))} . 

Furthermore, when Codex generated extra, unspecified code, learners had to evaluate whether to delete, keep, or modify the extra code. We analyzed 166 of such instances and found 103 instances (62\%) in which learners followed the expected behavior and handled the extra code correctly, while in 50 instances (30\%) learners mistakenly retained the extra incorrect code, or in 12 instances (7\%) learners deleted the extra useful code.

In 72 cases, Codex generated incorrect code that should have been erased (e.g., when it repeated existing incorrect code). Learners immediately deleted the extra non-useful part of the code in 62\% (n=45) cases, while keeping the incorrect part in 38\% (n=27) cases. In extreme cases like when Codex generated only comments for $P_{12}$, they executed the code which did not produce any output, but then proceeded and submitted the comments. In a similar situation, $P_{14}$ had written six lines of code but faced a logical issue they could not fix. They then prompted Codex with the copy of the entire task description. However, Codex repeated their existing incorrect code. Despite this, $P_{14}$ replaced their original code with the incorrectly repeated code and submitted it without testing instead of deleting the incorrect AI-generated code.

\subsubsection{Using AI-Generated Code as Example}

\begin{figure*}
    \centering
    \includegraphics[width=\textwidth]{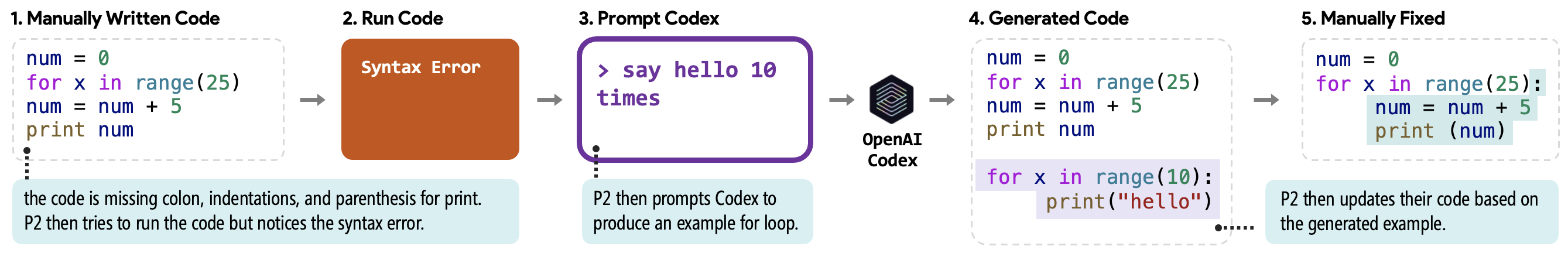}
    \caption{An example of using AI-generated code as an example to fix syntax error with writing loops.}
    \Description{An example of using AI-generated code as an example to fix syntax error with writing loops.}
    \label{fig:use_as_example}
\end{figure*}

We identified 26 instances in which learners used the AI-generated code as an example to fix their existing code or write new code similar to the generated code, instead of replacing their code with the newly generated code. For example, $P_6$ was struggling with concatenating a string and an integer, so they prompted Codex with \textit{“message = "num is: " plus variable num”} which generated \inlinecode{message = "num is: " + str(num)}. However, $P_6$ did not simply replace their incorrect code with the AI-generated code, but instead carefully compared the two codes and fixed their own code manually. Similarly, $P_3$ was trying to write an if-else conditional for the first time, however, not knowing the proper syntax, they encountered many issues. Therefore, they prompted Codex with “if else” which generated a sample code for them to fix their code. Similarly, when  $P_2$ was trying to write a for loop to accumulate a number by five for 25 times, they forgot about the colon and indentation which caused an error. To receive help, $P_2$ prompted Codex with \textit{“say hello 10 times”} and then fixed their code with the AI-generated code as an example (Figure \ref{fig:use_as_example}).

\subsection{RQ2: Effect of AI Code Generator Coding Approaches }

In previous sections, we presented results for each Codex usage. Now we shift our focus to how novices used Codex alongside manual programming approaches. Our analysis identified four distinct coding approaches used by learners to incorporate Codex into their programming practice: \textit{\textbf{AI Single Prompt}}, \textbf{\textit{AI Step-by-Step}}, \textbf{\textit{Hybrid}}, and \textbf{\textit{Manual}}. For each coding approach, we provide a description of the practice, descriptive statistics related to its usage, correlation with code-authoring and code-modification scores during the training phase. See Table \ref{tab:usage_pattern_auth_mod_scores} for a summary of statistics for each coding approach. We observed a limited number of tasks (n=48, 3.5\%) where learners switched approaches, for which we selected their final approach for our quantitative analyses.


\begin{figure*}
    \centering
    \includegraphics[width=\textwidth]{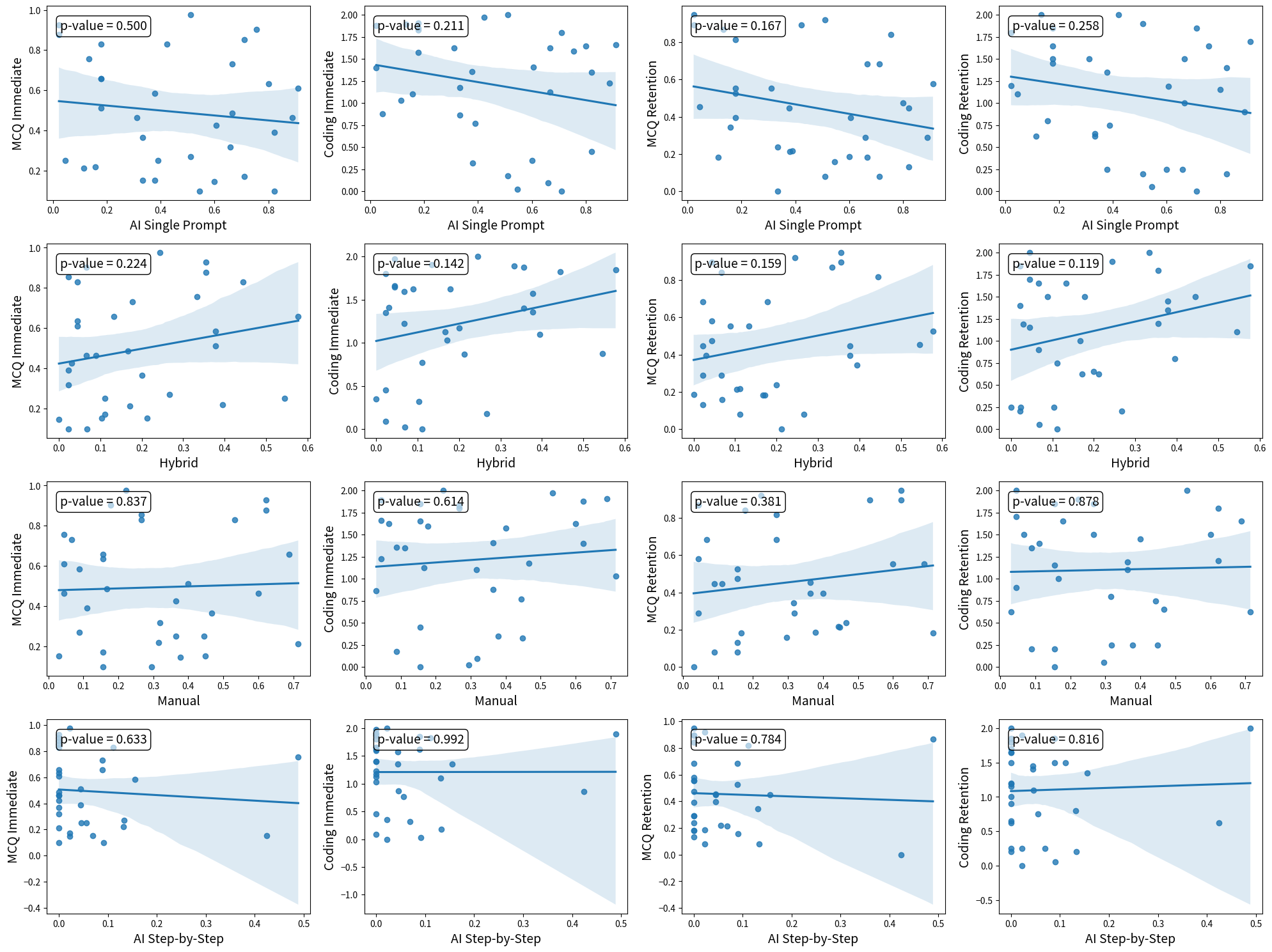}
    \caption{Each row represents correlations between the utilization of a coding approach and each of the four evaluation post-test scores.}
    \Description{Correlations between utilization of each coding approach and and evaluation test items.}
    \label{fig:coding_util_corr}
\end{figure*}

\subsubsection{AI Code Generator Coding Approaches}

\textbf{\textit{AI Single Prompt}}: The most frequently used coding approach (n=628, 46\%) was using a single prompt to generate the entire solution (Table \ref{tab:usage_pattern_auth_mod_scores}, Row 1). On 400 tasks, learners prompted Codex with a copy of the task description and submitted the AI-generated code with no manual coding (17 tasks were submitted without testing). Additionally, on 48 tasks, learners wrote the prompt message by themselves using their own words and understanding of the task.

\textbf{\textit{AI Step-by-Step}}: In this coding approach, which was used in 82 tasks (6\%), the main body of the submitted code was generated by Codex through multiple and consecutive Codex usages for different parts of the task (Table \ref{tab:usage_pattern_auth_mod_scores}, Row 2). Learners identified multiple subgoals based on the task description or simply broke the task description by its sentences and then prompted Codex with each of the parts.

\textbf{\textit{Hybrid}}: In this coding approach that was used for completing 252 tasks (19\%), a few subgoals of the submitted task were generated by Codex while the rest were written manually (Table \ref{tab:usage_pattern_auth_mod_scores}, Row 3). On 57 tasks, learners used Codex after manual coding only to debug their code and fix an encountered issue.

\textbf{\textit{Manual}}: In this coding approach that was used for completing 398 tasks (29\%), the final solution submitted by the learners was 100\% self-written (Table \ref{tab:usage_pattern_auth_mod_scores}, Row 4). In a special variant of this coding approach that was used in 42 tasks, learners used Codex, but did not use the AI-generated code due to its low-quality.

\begin{table*}[t]
\centering
\caption{Usage count and task correctness score averaged across all tasks in each coding approach.}
\label{tab:usage_pattern_auth_mod_scores}
\begin{tabular}{l l l l}
    \toprule
    Coding Approach & Usage & Authoring Score (\%) & Modifying Score (\%) \\
    \midrule
    \textbf{\textit{AI Single Prompt}} & 630 & \textit{M}=96\%, SD=16\% & \textit{M}=62\%, SD=45\% \\
    \textbf{\textit{AI Step-by-Step}} & 82 & \textit{M}=76\%, SD=37\% & \textit{M}=55\%, SD=47\% \\
    \textbf{\textit{Hybrid}} & 252 & \textit{M}=77\%, SD=34\% & \textit{M}=73\%, SD=39\% \\
    \textbf{\textit{Manual}} & 397 & \textit{M}=63\%, SD=43\% & \textit{M}=73\%, SD=39\% \\
  \bottomrule
\end{tabular}
\end{table*}

\subsubsection{Effect of Coding Approaches on Individual Learning Outcomes}

For every student, we calculated the utilization rate of each coding approach as the number of tasks they used a particular approach divided by the total number of tasks they completed during the training phase. We then calculated a series of Linear Regressions to investigate the relationship between the utilization rate of each coding approach and the immediate and retention post-test scores (on coding tasks and conceptual-MCQs). Our results presented in Figure \ref{fig:coding_util_corr} show the relationship between the utilization of each coding approach and performance on post-test evaluation tests. Although our analysis did not reveal any significant relationships between any of the coding approaches and the post-test evaluation tests, it shows consistently \textbf{positive} trends between the \textit{Hybrid} approach and post-test scores as well as consistently \textbf{negative} trends between the \textit{AI Single Prompt} approach and post-test evaluation scores. A possible interpretation of these findings suggests that individuals who actively switched between manual and AI-assisted coding for different parts of a task, demonstrated higher learning outcomes. However, future, more in-depth studies are required to explore the effect of coding approaches on individual learning outcomes.

%% file: sections/5-discussion.tex
\section{Discussion}

\subsection{Signs of Over-Reliance and Self-Regulation}

We have identified multiple signs of self-regulation on AI code generators, where learners executed cognitive control during a learning task \cite{prather2020we}. For example, learners actively added code to verify AI-generated code, or tinkered with the AI-generated code to understand the underlying concepts. Additionally, when learners temporarily removed a code that they deemed unnecessary and tested the code to see how it contributed to the program, they were actively engaged in the learning activity and showed signs of self-regulation. Similarly, when learners used AI-generated code as an example to fix their own code, they realized the importance of practice in learning.

However, we also discovered signs of over-reliance \cite{becker2023programming}. Research in Human-AI collaboration has reported that humans tend to over-trust and rely too heavily on AI agents \cite{passi2022overreliance}, even when they are incorrect \cite{buccinca2021trust, cao2022understanding}. Furthermore, as a learning support tool, such over-reliance can also connect with ineffective help-seeking behaviors in Interactive Learning Environments (ILE). ILEs are computer-based instructional systems that help beginners learn through task-based environments and support \cite{aleven2006toward, marwan2020unproductive}. When students use the on-demand system help, they often rely too heavily on the system's bottom-out hints that provide the correct answer with little to no explanation \cite{aleven2006toward}.

In our analysis, a common sign of over-reliance was the frequent use of the \textit{AI Single Prompt} coding approach, in which learners generated the entire solution using a single prompt. An extreme case of such over-reliance was when learners employed this approach at the beginning, prompted Codex with a copy of the task description, and submitted the generated code without any further editing. This is similar to students’ heavy reliance on bottom-out hints in ILEs which are more likely to harm their learning gains \cite{baker2004detecting, marwan2020unproductive}. For example, such over-reliance could result in fake progress during training and difficulties in applying basic concepts later in the training. Another sign of over-reliance was prompting Codex for code similar to existing code. This is related to self-regulation behaviors in help-seeking during problem-solving. Furthermore, this behavior is against the requirement of “reinvestment of received help”, where students are expected to reuse the received help in analogous tasks \cite{puustinen1998help}. Other signs of over-reliance were related to students’ potential over-trust of AI systems. Compared to traditional computer-based programming learning environments that provided correct support \cite{price2017hint}, AI systems might produce low-quality code that does not lead to the correct solution. Two representative signs of this over-reliance are accepting AI-generated code without verification and misplacement of AI-generated code. This indicates that some students might believe that AI code generators are fully capable of providing flawless code and they can insert their output into the intended location. Therefore, they assumed that no further action or verification was required by themselves.

Overall, as AI code generators are becoming more prevalent, effective, and accessible through tools like ChatGPT and Copilot, it is important to promote behaviors that contribute to higher learning outcomes while increasing awareness of the dangers of such tools. over-reliance causes learners to process AI information superficially, rather than critically engaging with it using their own knowledge \cite{gajos2022people}. Therefore, CS-Ed researchers and educators need to be aware of students’ potential misbehavior and integrate best practices of using AI code generators as part of the lesson plan.

\subsection{Implications for Designers}

Although prior research \cite{denny2023conversing, finnie2022robots, finnie2023my} investigated the impact of prompts on the quality of AI-generated code, those studies were mainly conducted offline, without user studies. In contrast, this work provides analyses of how young learners craft prompts, in terms of accuracy, language, and relationship to code. Our analysis also found that poorly crafted prompts led to low-quality code that learners struggled to work with. Therefore, future AI-generation tools could be inspired by \textit{Grounded Abstraction Matching} \cite{liu2023wants} and support novices in crafting prompts that have a one-to-one relationship with generated code without any missing details, as both a learning activity and programming support tool.

Moreover, to encourage the self-regulated use of AI code generators, such as tinkering, future tools could offer a sandbox for learners to experiment with the prompt until they achieve the desired output through iterative testing and verification. This could increase playfulness as learners may feel less attached to the generated code and be more willing to tinker with it. The sandbox could also be designed in a way to promote tinkering, such as incorporating visual cues that differentiate editable values from fixed syntactical parts of the generated code. These visual cues could be designed in a way to promotes playful tinkering. In addition, properly decomposing problems can support program development processes \cite{kwon2019exploring}. However, we found that some students were not effective at task decomposition. Specifically, some students resorted to dividing the task description sentence by sentence for each prompt. Therefore, future design should consider incorporating guided planning components \cite{jin2014evaluation} or decomposition charts to train students how to properly decompose a task.  LLMs can also help with the decomposition of a task. Future AI code generators targeted at novices could break down a natural language programming prompt into sub-tasks and display them as a Parsons problem. This way not only the solution is not revealed, but learners are required to actively engage in a puzzle activity before writing code.

%% file: sections/6-limitations.tex
\section{Limitations}

The results presented are based on the participants' uninterrupted interaction with the AI code generator, and do not consider their thought processes and motivations while using Codex. This limitation restricts the depth of our qualitative analysis. Therefore, to fully comprehend the exact motivations behind each Codex usage, what each prompt message meant to the participants, their intentions behind modifying the AI-generated code, and how they verified the code, future introspective and retrospective think-aloud studies are necessary. In addition, our analysis involved only 33 young learners in an online and informal study setting. Other behaviors and usage scenarios may be uncovered with different age groups, as well as in more formal classroom settings like a K-12 or undergraduate programming course. Furthermore, certain behaviors, such as placing the AI-generated code incorrectly, are specific to the way Coding Steps was designed, and therefore may not generalize to other AI code generators. 



%% file: sections/7-conclusion.tex
\section{Conclusion}

With the rapid access learners have to Large Language Models (LLMs) through tools like ChatGPT and Github Copilot (based on OpenAI Codex), there are rising concerns about proper learning, over-reliance, and plagiarism. However, these tools are here to stay; therefore, it is necessary to adapt the future of curriculum and tool development based on the most effective strategies and practices that learners use with AI code generators.

This work provides initial insights into how learners use AI code generators, including common usage patterns, the types of prompts they use, and how learners verify and use AI-generated code. Our detailed thematic analysis provide evidence of various signs of over-reliance and self-regulation in learners when interacting with LLM-based code generators. Some learners dealt with AI-generated code properly even when using Codex in "\textit{autonomous}" modes like the \textit{AI Single Prompt} approach. They manually added code to verify it, and even tinkered with it to better understand the generated code, however, some participants had difficulty integrating the AI-generated code into their solution. In addition, students not only used the original task description to ask for code, but they also crafted prompts with varied language and clarity levels, which sometimes affected the quality of the generated code. Furthermore, we identified four general coding approaches when learners write code with AI code generators: \textit{AI Single Prompt}, \textit{AI Step-by-Step}, \textit{Hybrid}, and \textit{Manual}. We also analyzed the effect of different coding approaches on learning outcomes tested a week later. Our analysis shows consistently \textbf{negative} trends between the utilization of the \textit{AI Single Prompt} approach and post-test evaluation scores as well as consistently \textbf{positive} trends between the utilization of the \textit{Hybrid} approach and post-test evaluation scores.

%% file: main.bbl

\begin{thebibliography}{54}


\ifx \showCODEN    \undefined \def \showCODEN     #1{\unskip}     \fi
\ifx \showDOI      \undefined \def \showDOI       #1{#1}\fi
\ifx \showISBNx    \undefined \def \showISBNx     #1{\unskip}     \fi
\ifx \showISBNxiii \undefined \def \showISBNxiii  #1{\unskip}     \fi
\ifx \showISSN     \undefined \def \showISSN      #1{\unskip}     \fi
\ifx \showLCCN     \undefined \def \showLCCN      #1{\unskip}     \fi
\ifx \shownote     \undefined \def \shownote      #1{#1}          \fi
\ifx \showarticletitle \undefined \def \showarticletitle #1{#1}   \fi
\ifx \showURL      \undefined \def \showURL       {\relax}        \fi
\providecommand\bibfield[2]{#2}
\providecommand\bibinfo[2]{#2}
\providecommand\natexlab[1]{#1}
\providecommand\showeprint[2][]{arXiv:#2}

\bibitem[Aleven et~al\mbox{.}(2006)]%
        {aleven2006toward}
\bibfield{author}{\bibinfo{person}{Vincent Aleven}, \bibinfo{person}{Bruce
  McLaren}, \bibinfo{person}{Ido Roll}, {and} \bibinfo{person}{Kenneth
  Koedinger}.} \bibinfo{year}{2006}\natexlab{}.
\newblock \showarticletitle{Toward meta-cognitive tutoring: A model of help
  seeking with a Cognitive Tutor}.
\newblock \bibinfo{journal}{\emph{International Journal of Artificial
  Intelligence in Education}} \bibinfo{volume}{16}, \bibinfo{number}{2}
  (\bibinfo{year}{2006}), \bibinfo{pages}{101--128}.
\newblock


\bibitem[Austin et~al\mbox{.}(2021)]%
        {austin2021program}
\bibfield{author}{\bibinfo{person}{Jacob Austin}, \bibinfo{person}{Augustus
  Odena}, \bibinfo{person}{Maxwell Nye}, \bibinfo{person}{Maarten Bosma},
  \bibinfo{person}{Henryk Michalewski}, \bibinfo{person}{David Dohan},
  \bibinfo{person}{Ellen Jiang}, \bibinfo{person}{Carrie Cai},
  \bibinfo{person}{Michael Terry}, \bibinfo{person}{Quoc Le}, {et~al\mbox{.}}}
  \bibinfo{year}{2021}\natexlab{}.
\newblock \showarticletitle{Program synthesis with large language models}.
\newblock \bibinfo{journal}{\emph{arXiv preprint arXiv:2108.07732}}
  (\bibinfo{year}{2021}).
\newblock


\bibitem[Baker et~al\mbox{.}(2004)]%
        {baker2004detecting}
\bibfield{author}{\bibinfo{person}{Ryan~Shaun Baker}, \bibinfo{person}{Albert~T
  Corbett}, {and} \bibinfo{person}{Kenneth~R Koedinger}.}
  \bibinfo{year}{2004}\natexlab{}.
\newblock \showarticletitle{Detecting student misuse of intelligent tutoring
  systems}. In \bibinfo{booktitle}{\emph{Intelligent Tutoring Systems: 7th
  International Conference, ITS 2004, Macei{\'o}, Alagoas, Brazil, August
  30-September 3, 2004. Proceedings 7}}. Springer, \bibinfo{pages}{531--540}.
\newblock


\bibitem[Barke et~al\mbox{.}(2023)]%
        {barke2023grounded}
\bibfield{author}{\bibinfo{person}{Shraddha Barke}, \bibinfo{person}{Michael~B
  James}, {and} \bibinfo{person}{Nadia Polikarpova}.}
  \bibinfo{year}{2023}\natexlab{}.
\newblock \showarticletitle{Grounded copilot: How programmers interact with
  code-generating models}.
\newblock \bibinfo{journal}{\emph{Proceedings of the ACM on Programming
  Languages}} \bibinfo{volume}{7}, \bibinfo{number}{OOPSLA1}
  (\bibinfo{year}{2023}), \bibinfo{pages}{85--111}.
\newblock


\bibitem[Becker et~al\mbox{.}(2023)]%
        {becker2023programming}
\bibfield{author}{\bibinfo{person}{Brett~A Becker}, \bibinfo{person}{Paul
  Denny}, \bibinfo{person}{James Finnie-Ansley}, \bibinfo{person}{Andrew
  Luxton-Reilly}, \bibinfo{person}{James Prather}, {and}
  \bibinfo{person}{Eddie~Antonio Santos}.} \bibinfo{year}{2023}\natexlab{}.
\newblock \showarticletitle{Programming Is Hard-Or at Least It Used to Be:
  Educational Opportunities and Challenges of AI Code Generation}. In
  \bibinfo{booktitle}{\emph{Proceedings of the 54th ACM Technical Symposium on
  Computer Science Education V. 1}}. \bibinfo{pages}{500--506}.
\newblock


\bibitem[Bernard et~al\mbox{.}(2016)]%
        {bernard2016analyzing}
\bibfield{author}{\bibinfo{person}{H~Russell Bernard}, \bibinfo{person}{Amber
  Wutich}, {and} \bibinfo{person}{Gery~W Ryan}.}
  \bibinfo{year}{2016}\natexlab{}.
\newblock \bibinfo{booktitle}{\emph{Analyzing qualitative data: Systematic
  approaches}}.
\newblock \bibinfo{publisher}{SAGE publications}.
\newblock


\bibitem[Bingham and Witkowsky(2021)]%
        {bingham2021deductive}
\bibfield{author}{\bibinfo{person}{Andrea~J Bingham} {and}
  \bibinfo{person}{Patricia Witkowsky}.} \bibinfo{year}{2021}\natexlab{}.
\newblock \showarticletitle{Deductive and inductive approaches to qualitative
  data analysis}.
\newblock \bibinfo{journal}{\emph{Analyzing and interpreting qualitative data:
  After the interview}} (\bibinfo{year}{2021}), \bibinfo{pages}{133--146}.
\newblock


\bibitem[Brandt et~al\mbox{.}(2009)]%
        {brandt2009two}
\bibfield{author}{\bibinfo{person}{Joel Brandt}, \bibinfo{person}{Philip~J
  Guo}, \bibinfo{person}{Joel Lewenstein}, \bibinfo{person}{Mira Dontcheva},
  {and} \bibinfo{person}{Scott~R Klemmer}.} \bibinfo{year}{2009}\natexlab{}.
\newblock \showarticletitle{Two studies of opportunistic programming:
  interleaving web foraging, learning, and writing code}. In
  \bibinfo{booktitle}{\emph{Proceedings of the SIGCHI Conference on Human
  Factors in Computing Systems}}. \bibinfo{pages}{1589--1598}.
\newblock


\bibitem[Bruner et~al\mbox{.}(1966)]%
        {bruner1966toward}
\bibfield{author}{\bibinfo{person}{Jerome~Seymour Bruner} {et~al\mbox{.}}}
  \bibinfo{year}{1966}\natexlab{}.
\newblock \bibinfo{booktitle}{\emph{Toward a theory of instruction}}.
  Vol.~\bibinfo{volume}{59}.
\newblock \bibinfo{publisher}{Harvard University Press}.
\newblock


\bibitem[Bu{\c{c}}inca et~al\mbox{.}(2021)]%
        {buccinca2021trust}
\bibfield{author}{\bibinfo{person}{Zana Bu{\c{c}}inca},
  \bibinfo{person}{Maja~Barbara Malaya}, {and} \bibinfo{person}{Krzysztof~Z
  Gajos}.} \bibinfo{year}{2021}\natexlab{}.
\newblock \showarticletitle{To trust or to think: cognitive forcing functions
  can reduce overreliance on AI in AI-assisted decision-making}.
\newblock \bibinfo{journal}{\emph{Proceedings of the ACM on Human-Computer
  Interaction}} \bibinfo{volume}{5}, \bibinfo{number}{CSCW1}
  (\bibinfo{year}{2021}), \bibinfo{pages}{1--21}.
\newblock


\bibitem[Cao and Huang(2022)]%
        {cao2022understanding}
\bibfield{author}{\bibinfo{person}{Shiye Cao} {and} \bibinfo{person}{Chien-Ming
  Huang}.} \bibinfo{year}{2022}\natexlab{}.
\newblock \showarticletitle{Understanding User Reliance on AI in Assisted
  Decision-Making}.
\newblock \bibinfo{journal}{\emph{Proceedings of the ACM on Human-Computer
  Interaction}} \bibinfo{volume}{6}, \bibinfo{number}{CSCW2}
  (\bibinfo{year}{2022}), \bibinfo{pages}{1--23}.
\newblock


\bibitem[Chen et~al\mbox{.}(2021)]%
        {chen2021evaluating}
\bibfield{author}{\bibinfo{person}{Mark Chen}, \bibinfo{person}{Jerry Tworek},
  \bibinfo{person}{Heewoo Jun}, \bibinfo{person}{Qiming Yuan},
  \bibinfo{person}{Henrique Ponde de~Oliveira Pinto}, \bibinfo{person}{Jared
  Kaplan}, \bibinfo{person}{Harri Edwards}, \bibinfo{person}{Yuri Burda},
  \bibinfo{person}{Nicholas Joseph}, \bibinfo{person}{Greg Brockman},
  {et~al\mbox{.}}} \bibinfo{year}{2021}\natexlab{}.
\newblock \showarticletitle{Evaluating large language models trained on code}.
\newblock \bibinfo{journal}{\emph{arXiv preprint arXiv:2107.03374}}
  (\bibinfo{year}{2021}).
\newblock


\bibitem[Chowdhery et~al\mbox{.}(2022)]%
        {chowdhery2022palm}
\bibfield{author}{\bibinfo{person}{Aakanksha Chowdhery},
  \bibinfo{person}{Sharan Narang}, \bibinfo{person}{Jacob Devlin},
  \bibinfo{person}{Maarten Bosma}, \bibinfo{person}{Gaurav Mishra},
  \bibinfo{person}{Adam Roberts}, \bibinfo{person}{Paul Barham},
  \bibinfo{person}{Hyung~Won Chung}, \bibinfo{person}{Charles Sutton},
  \bibinfo{person}{Sebastian Gehrmann}, {et~al\mbox{.}}}
  \bibinfo{year}{2022}\natexlab{}.
\newblock \showarticletitle{Palm: Scaling language modeling with pathways}.
\newblock \bibinfo{journal}{\emph{arXiv preprint arXiv:2204.02311}}
  (\bibinfo{year}{2022}).
\newblock


\bibitem[Cunningham et~al\mbox{.}(2021)]%
        {cunningham2021avoiding}
\bibfield{author}{\bibinfo{person}{Kathryn Cunningham},
  \bibinfo{person}{Barbara~J Ericson}, \bibinfo{person}{Rahul
  Agrawal~Bejarano}, {and} \bibinfo{person}{Mark Guzdial}.}
  \bibinfo{year}{2021}\natexlab{}.
\newblock \showarticletitle{Avoiding the Turing tarpit: Learning conversational
  programming by starting from code’s purpose}. In
  \bibinfo{booktitle}{\emph{Proceedings of the 2021 CHI Conference on Human
  Factors in Computing Systems}}. \bibinfo{pages}{1--15}.
\newblock


\bibitem[Denny et~al\mbox{.}(2023)]%
        {denny2023conversing}
\bibfield{author}{\bibinfo{person}{Paul Denny}, \bibinfo{person}{Viraj Kumar},
  {and} \bibinfo{person}{Nasser Giacaman}.} \bibinfo{year}{2023}\natexlab{}.
\newblock \showarticletitle{Conversing with copilot: Exploring prompt
  engineering for solving cs1 problems using natural language}. In
  \bibinfo{booktitle}{\emph{Proceedings of the 54th ACM Technical Symposium on
  Computer Science Education V. 1}}. \bibinfo{pages}{1136--1142}.
\newblock


\bibitem[Denny et~al\mbox{.}(2022)]%
        {denny2022robosourcing}
\bibfield{author}{\bibinfo{person}{Paul Denny}, \bibinfo{person}{Sami Sarsa},
  \bibinfo{person}{Arto Hellas}, {and} \bibinfo{person}{Juho Leinonen}.}
  \bibinfo{year}{2022}\natexlab{}.
\newblock \showarticletitle{Robosourcing Educational Resources--Leveraging
  Large Language Models for Learnersourcing}.
\newblock \bibinfo{journal}{\emph{arXiv preprint arXiv:2211.04715}}
  (\bibinfo{year}{2022}).
\newblock


\bibitem[Finnie-Ansley et~al\mbox{.}(2022)]%
        {finnie2022robots}
\bibfield{author}{\bibinfo{person}{James Finnie-Ansley}, \bibinfo{person}{Paul
  Denny}, \bibinfo{person}{Brett~A Becker}, \bibinfo{person}{Andrew
  Luxton-Reilly}, {and} \bibinfo{person}{James Prather}.}
  \bibinfo{year}{2022}\natexlab{}.
\newblock \showarticletitle{The robots are coming: Exploring the implications
  of openai codex on introductory programming}. In
  \bibinfo{booktitle}{\emph{Australasian Computing Education Conference}}.
  \bibinfo{pages}{10--19}.
\newblock


\bibitem[Finnie-Ansley et~al\mbox{.}(2023)]%
        {finnie2023my}
\bibfield{author}{\bibinfo{person}{James Finnie-Ansley}, \bibinfo{person}{Paul
  Denny}, \bibinfo{person}{Andrew Luxton-Reilly},
  \bibinfo{person}{Eddie~Antonio Santos}, \bibinfo{person}{James Prather},
  {and} \bibinfo{person}{Brett~A Becker}.} \bibinfo{year}{2023}\natexlab{}.
\newblock \showarticletitle{My AI Wants to Know if This Will Be on the Exam:
  Testing OpenAI’s Codex on CS2 Programming Exercises}. In
  \bibinfo{booktitle}{\emph{Proceedings of the 25th Australasian Computing
  Education Conference}}. \bibinfo{pages}{97--104}.
\newblock


\bibitem[Gajos and Mamykina(2022)]%
        {gajos2022people}
\bibfield{author}{\bibinfo{person}{Krzysztof~Z Gajos} {and}
  \bibinfo{person}{Lena Mamykina}.} \bibinfo{year}{2022}\natexlab{}.
\newblock \showarticletitle{Do people engage cognitively with ai? impact of ai
  assistance on incidental learning}. In \bibinfo{booktitle}{\emph{27th
  International Conference on Intelligent User Interfaces}}.
  \bibinfo{pages}{794--806}.
\newblock


\bibitem[{Github}(2022)]%
        {copilot}
\bibfield{author}{\bibinfo{person}{{Github}}.} \bibinfo{year}{2022}\natexlab{}.
\newblock \bibinfo{title}{{Copilot: Your AI pair programmer}}.
\newblock \bibinfo{howpublished}{{\url{https://github.com/features/copilot}}}.
\newblock
\newblock
\shownote{[Online; accessed 9-September-2022]}.


\bibitem[Hou et~al\mbox{.}(2022)]%
        {hou2022using}
\bibfield{author}{\bibinfo{person}{Xinying Hou}, \bibinfo{person}{Barbara~Jane
  Ericson}, {and} \bibinfo{person}{Xu Wang}.} \bibinfo{year}{2022}\natexlab{}.
\newblock \showarticletitle{Using Adaptive Parsons Problems to Scaffold
  Write-Code Problems}. In \bibinfo{booktitle}{\emph{Proceedings of the 2022
  ACM Conference on International Computing Education Research-Volume 1}}.
  \bibinfo{pages}{15--26}.
\newblock


\bibitem[Ichinco and Kelleher(2015)]%
        {ichinco2015exploring}
\bibfield{author}{\bibinfo{person}{Michelle Ichinco} {and}
  \bibinfo{person}{Caitlin Kelleher}.} \bibinfo{year}{2015}\natexlab{}.
\newblock \showarticletitle{Exploring novice programmer example use}. In
  \bibinfo{booktitle}{\emph{2015 IEEE Symposium on Visual Languages and
  Human-Centric Computing (VL/HCC)}}. IEEE, \bibinfo{pages}{63--71}.
\newblock


\bibitem[Jain et~al\mbox{.}(2022)]%
        {jain2022jigsaw}
\bibfield{author}{\bibinfo{person}{Naman Jain}, \bibinfo{person}{Skanda
  Vaidyanath}, \bibinfo{person}{Arun Iyer}, \bibinfo{person}{Nagarajan
  Natarajan}, \bibinfo{person}{Suresh Parthasarathy}, \bibinfo{person}{Sriram
  Rajamani}, {and} \bibinfo{person}{Rahul Sharma}.}
  \bibinfo{year}{2022}\natexlab{}.
\newblock \showarticletitle{Jigsaw: Large language models meet program
  synthesis}. In \bibinfo{booktitle}{\emph{Proceedings of the 44th
  International Conference on Software Engineering}}.
  \bibinfo{pages}{1219--1231}.
\newblock


\bibitem[Jayagopal et~al\mbox{.}(2022)]%
        {jayagopal2022exploring}
\bibfield{author}{\bibinfo{person}{Dhanya Jayagopal}, \bibinfo{person}{Justin
  Lubin}, {and} \bibinfo{person}{Sarah~E Chasins}.}
  \bibinfo{year}{2022}\natexlab{}.
\newblock \showarticletitle{Exploring the Learnability of Program Synthesizers
  by Novice Programmers}. In \bibinfo{booktitle}{\emph{Proceedings of the 35th
  Annual ACM Symposium on User Interface Software and Technology}}.
  \bibinfo{pages}{1--15}.
\newblock


\bibitem[Jiang et~al\mbox{.}(2022)]%
        {jiang2022discovering}
\bibfield{author}{\bibinfo{person}{Ellen Jiang}, \bibinfo{person}{Edwin Toh},
  \bibinfo{person}{Alejandra Molina}, \bibinfo{person}{Kristen Olson},
  \bibinfo{person}{Claire Kayacik}, \bibinfo{person}{Aaron Donsbach},
  \bibinfo{person}{Carrie~J Cai}, {and} \bibinfo{person}{Michael Terry}.}
  \bibinfo{year}{2022}\natexlab{}.
\newblock \showarticletitle{Discovering the syntax and strategies of natural
  language programming with generative language models}. In
  \bibinfo{booktitle}{\emph{Proceedings of the 2022 CHI Conference on Human
  Factors in Computing Systems}}. \bibinfo{pages}{1--19}.
\newblock


\bibitem[Jin et~al\mbox{.}(2014)]%
        {jin2014evaluation}
\bibfield{author}{\bibinfo{person}{Wei Jin}, \bibinfo{person}{Albert Corbett},
  \bibinfo{person}{Will Lloyd}, \bibinfo{person}{Lewis Baumstark}, {and}
  \bibinfo{person}{Christine Rolka}.} \bibinfo{year}{2014}\natexlab{}.
\newblock \showarticletitle{Evaluation of guided-planning and assisted-coding
  with task relevant dynamic hinting}. In \bibinfo{booktitle}{\emph{Intelligent
  Tutoring Systems: 12th International Conference, ITS 2014, Honolulu, HI, USA,
  June 5-9, 2014. Proceedings 12}}. Springer, \bibinfo{pages}{318--328}.
\newblock


\bibitem[Kasneci et~al\mbox{.}(2023)]%
        {kasneci2023chatgpt}
\bibfield{author}{\bibinfo{person}{Enkelejda Kasneci}, \bibinfo{person}{Kathrin
  Se{\ss}ler}, \bibinfo{person}{Stefan K{\"u}chemann}, \bibinfo{person}{Maria
  Bannert}, \bibinfo{person}{Daryna Dementieva}, \bibinfo{person}{Frank
  Fischer}, \bibinfo{person}{Urs Gasser}, \bibinfo{person}{Georg Groh},
  \bibinfo{person}{Stephan G{\"u}nnemann}, \bibinfo{person}{Eyke
  H{\"u}llermeier}, {et~al\mbox{.}}} \bibinfo{year}{2023}\natexlab{}.
\newblock \showarticletitle{ChatGPT for good? On opportunities and challenges
  of large language models for education}.
\newblock \bibinfo{journal}{\emph{Learning and Individual Differences}}
  \bibinfo{volume}{103} (\bibinfo{year}{2023}), \bibinfo{pages}{102274}.
\newblock


\bibitem[Kazemitabaar et~al\mbox{.}(2023)]%
        {kazemitabaar2023studying}
\bibfield{author}{\bibinfo{person}{Majeed Kazemitabaar},
  \bibinfo{person}{Justin Chow}, \bibinfo{person}{Carl Ka~To Ma},
  \bibinfo{person}{Barbara~J Ericson}, \bibinfo{person}{David Weintrop}, {and}
  \bibinfo{person}{Tovi Grossman}.} \bibinfo{year}{2023}\natexlab{}.
\newblock \showarticletitle{Studying the effect of AI Code Generators on
  Supporting Novice Learners in Introductory Programming}. In
  \bibinfo{booktitle}{\emph{Proceedings of the 2023 CHI Conference on Human
  Factors in Computing Systems}}. \bibinfo{pages}{1--23}.
\newblock


\bibitem[Keuning et~al\mbox{.}(2018)]%
        {keuning2018systematic}
\bibfield{author}{\bibinfo{person}{Hieke Keuning}, \bibinfo{person}{Johan
  Jeuring}, {and} \bibinfo{person}{Bastiaan Heeren}.}
  \bibinfo{year}{2018}\natexlab{}.
\newblock \showarticletitle{A systematic literature review of automated
  feedback generation for programming exercises}.
\newblock \bibinfo{journal}{\emph{ACM Transactions on Computing Education
  (TOCE)}} \bibinfo{volume}{19}, \bibinfo{number}{1} (\bibinfo{year}{2018}),
  \bibinfo{pages}{1--43}.
\newblock


\bibitem[Kwon and Cheon(2019)]%
        {kwon2019exploring}
\bibfield{author}{\bibinfo{person}{Kyungbin Kwon} {and}
  \bibinfo{person}{Jongpil Cheon}.} \bibinfo{year}{2019}\natexlab{}.
\newblock \showarticletitle{Exploring Problem Decomposition and Program
  Development through Block-Based Programs.}
\newblock \bibinfo{journal}{\emph{International Journal of Computer Science
  Education in Schools}} \bibinfo{volume}{3}, \bibinfo{number}{1}
  (\bibinfo{year}{2019}), \bibinfo{pages}{n1}.
\newblock


\bibitem[Leinonen et~al\mbox{.}(2023)]%
        {leinonen2023using}
\bibfield{author}{\bibinfo{person}{Juho Leinonen}, \bibinfo{person}{Arto
  Hellas}, \bibinfo{person}{Sami Sarsa}, \bibinfo{person}{Brent Reeves},
  \bibinfo{person}{Paul Denny}, \bibinfo{person}{James Prather}, {and}
  \bibinfo{person}{Brett~A Becker}.} \bibinfo{year}{2023}\natexlab{}.
\newblock \showarticletitle{Using large language models to enhance programming
  error messages}. In \bibinfo{booktitle}{\emph{Proceedings of the 54th ACM
  Technical Symposium on Computer Science Education V. 1}}.
  \bibinfo{pages}{563--569}.
\newblock


\bibitem[Li et~al\mbox{.}(2022)]%
        {li2022competition}
\bibfield{author}{\bibinfo{person}{Yujia Li}, \bibinfo{person}{David Choi},
  \bibinfo{person}{Junyoung Chung}, \bibinfo{person}{Nate Kushman},
  \bibinfo{person}{Julian Schrittwieser}, \bibinfo{person}{R{\'e}mi Leblond},
  \bibinfo{person}{Tom Eccles}, \bibinfo{person}{James Keeling},
  \bibinfo{person}{Felix Gimeno}, \bibinfo{person}{Agustin Dal~Lago},
  {et~al\mbox{.}}} \bibinfo{year}{2022}\natexlab{}.
\newblock \showarticletitle{Competition-level code generation with alphacode}.
\newblock \bibinfo{journal}{\emph{Science}} \bibinfo{volume}{378},
  \bibinfo{number}{6624} (\bibinfo{year}{2022}), \bibinfo{pages}{1092--1097}.
\newblock


\bibitem[Liu et~al\mbox{.}(2023)]%
        {liu2023wants}
\bibfield{author}{\bibinfo{person}{Michael~Xieyang Liu},
  \bibinfo{person}{Advait Sarkar}, \bibinfo{person}{Carina Negreanu},
  \bibinfo{person}{Benjamin Zorn}, \bibinfo{person}{Jack Williams},
  \bibinfo{person}{Neil Toronto}, {and} \bibinfo{person}{Andrew~D Gordon}.}
  \bibinfo{year}{2023}\natexlab{}.
\newblock \showarticletitle{“What It Wants Me To Say”: Bridging the
  Abstraction Gap Between End-User Programmers and Code-Generating Large
  Language Models}. In \bibinfo{booktitle}{\emph{Proceedings of the 2023 CHI
  Conference on Human Factors in Computing Systems}}. \bibinfo{pages}{1--31}.
\newblock


\bibitem[Marwan et~al\mbox{.}(2020)]%
        {marwan2020unproductive}
\bibfield{author}{\bibinfo{person}{Samiha Marwan}, \bibinfo{person}{Anay
  Dombe}, {and} \bibinfo{person}{Thomas~W Price}.}
  \bibinfo{year}{2020}\natexlab{}.
\newblock \showarticletitle{Unproductive help-seeking in programming: What it
  is and how to address it}. In \bibinfo{booktitle}{\emph{Proceedings of the
  2020 ACM conference on innovation and technology in computer science
  education}}. \bibinfo{pages}{54--60}.
\newblock


\bibitem[Miles and Huberman(1994)]%
        {miles1994qualitative}
\bibfield{author}{\bibinfo{person}{Matthew~B Miles} {and}
  \bibinfo{person}{A~Michael Huberman}.} \bibinfo{year}{1994}\natexlab{}.
\newblock \bibinfo{booktitle}{\emph{Qualitative data analysis: An expanded
  sourcebook}}.
\newblock \bibinfo{publisher}{sage}.
\newblock


\bibitem[Mozannar et~al\mbox{.}(2022)]%
        {mozannar2022reading}
\bibfield{author}{\bibinfo{person}{Hussein Mozannar}, \bibinfo{person}{Gagan
  Bansal}, \bibinfo{person}{Adam Fourney}, {and} \bibinfo{person}{Eric
  Horvitz}.} \bibinfo{year}{2022}\natexlab{}.
\newblock \showarticletitle{Reading Between the Lines: Modeling User Behavior
  and Costs in AI-Assisted Programming}.
\newblock \bibinfo{journal}{\emph{arXiv preprint arXiv:2210.14306}}
  (\bibinfo{year}{2022}).
\newblock


\bibitem[Neuendorf(2017)]%
        {neuendorf2017content}
\bibfield{author}{\bibinfo{person}{Kimberly~A Neuendorf}.}
  \bibinfo{year}{2017}\natexlab{}.
\newblock \bibinfo{booktitle}{\emph{The content analysis guidebook}}.
\newblock \bibinfo{publisher}{sage}.
\newblock


\bibitem[Passi and Vorvoreanu(2022)]%
        {passi2022overreliance}
\bibfield{author}{\bibinfo{person}{Samir Passi} {and} \bibinfo{person}{Mihaela
  Vorvoreanu}.} \bibinfo{year}{2022}\natexlab{}.
\newblock \showarticletitle{Overreliance on AI: Literature review}.
\newblock  (\bibinfo{year}{2022}).
\newblock


\bibitem[Phung et~al\mbox{.}(2023)]%
        {phung2023generating}
\bibfield{author}{\bibinfo{person}{Tung Phung}, \bibinfo{person}{Jos{\'e}
  Cambronero}, \bibinfo{person}{Sumit Gulwani}, \bibinfo{person}{Tobias Kohn},
  \bibinfo{person}{Rupak Majumdar}, \bibinfo{person}{Adish Singla}, {and}
  \bibinfo{person}{Gustavo Soares}.} \bibinfo{year}{2023}\natexlab{}.
\newblock \showarticletitle{Generating High-Precision Feedback for Programming
  Syntax Errors using Large Language Models}.
\newblock \bibinfo{journal}{\emph{arXiv preprint arXiv:2302.04662}}
  (\bibinfo{year}{2023}).
\newblock


\bibitem[Prather et~al\mbox{.}(2020)]%
        {prather2020we}
\bibfield{author}{\bibinfo{person}{James Prather}, \bibinfo{person}{Brett~A
  Becker}, \bibinfo{person}{Michelle Craig}, \bibinfo{person}{Paul Denny},
  \bibinfo{person}{Dastyni Loksa}, {and} \bibinfo{person}{Lauren Margulieux}.}
  \bibinfo{year}{2020}\natexlab{}.
\newblock \showarticletitle{What do we think we think we are doing?
  Metacognition and self-regulation in programming}. In
  \bibinfo{booktitle}{\emph{Proceedings of the 2020 ACM conference on
  international computing education research}}. \bibinfo{pages}{2--13}.
\newblock


\bibitem[Price et~al\mbox{.}(2017)]%
        {price2017hint}
\bibfield{author}{\bibinfo{person}{Thomas~W Price}, \bibinfo{person}{Rui Zhi},
  {and} \bibinfo{person}{Tiffany Barnes}.} \bibinfo{year}{2017}\natexlab{}.
\newblock \showarticletitle{Hint generation under uncertainty: The effect of
  hint quality on help-seeking behavior}. In
  \bibinfo{booktitle}{\emph{Artificial Intelligence in Education: 18th
  International Conference, AIED 2017, Wuhan, China, June 28--July 1, 2017,
  Proceedings 18}}. Springer, \bibinfo{pages}{311--322}.
\newblock


\bibitem[Puustinen(1998)]%
        {puustinen1998help}
\bibfield{author}{\bibinfo{person}{Minna Puustinen}.}
  \bibinfo{year}{1998}\natexlab{}.
\newblock \showarticletitle{Help-seeking behavior in a problem-solving
  situation: Development of self-regulation}.
\newblock \bibinfo{journal}{\emph{European Journal of Psychology of education}}
   \bibinfo{volume}{13} (\bibinfo{year}{1998}), \bibinfo{pages}{271--282}.
\newblock


\bibitem[Rivers and Koedinger(2017)]%
        {rivers2017data}
\bibfield{author}{\bibinfo{person}{Kelly Rivers} {and}
  \bibinfo{person}{Kenneth~R Koedinger}.} \bibinfo{year}{2017}\natexlab{}.
\newblock \showarticletitle{Data-driven hint generation in vast solution
  spaces: a self-improving python programming tutor}.
\newblock \bibinfo{journal}{\emph{International Journal of Artificial
  Intelligence in Education}}  \bibinfo{volume}{27} (\bibinfo{year}{2017}),
  \bibinfo{pages}{37--64}.
\newblock


\bibitem[Ross et~al\mbox{.}(2023)]%
        {ross2023programmer}
\bibfield{author}{\bibinfo{person}{Steven~I Ross}, \bibinfo{person}{Fernando
  Martinez}, \bibinfo{person}{Stephanie Houde}, \bibinfo{person}{Michael
  Muller}, {and} \bibinfo{person}{Justin~D Weisz}.}
  \bibinfo{year}{2023}\natexlab{}.
\newblock \showarticletitle{The programmer’s assistant: Conversational
  interaction with a large language model for software development}. In
  \bibinfo{booktitle}{\emph{Proceedings of the 28th International Conference on
  Intelligent User Interfaces}}. \bibinfo{pages}{491--514}.
\newblock


\bibitem[Sarkar et~al\mbox{.}(2022)]%
        {sarkar2022like}
\bibfield{author}{\bibinfo{person}{Advait Sarkar}, \bibinfo{person}{Andrew~D
  Gordon}, \bibinfo{person}{Carina Negreanu}, \bibinfo{person}{Christian
  Poelitz}, \bibinfo{person}{Sruti~Srinivasa Ragavan}, {and}
  \bibinfo{person}{Ben Zorn}.} \bibinfo{year}{2022}\natexlab{}.
\newblock \showarticletitle{What is it like to program with artificial
  intelligence?}
\newblock \bibinfo{journal}{\emph{arXiv preprint arXiv:2208.06213}}
  (\bibinfo{year}{2022}).
\newblock


\bibitem[Sarsa et~al\mbox{.}(2022)]%
        {sarsa2022automatic}
\bibfield{author}{\bibinfo{person}{Sami Sarsa}, \bibinfo{person}{Paul Denny},
  \bibinfo{person}{Arto Hellas}, {and} \bibinfo{person}{Juho Leinonen}.}
  \bibinfo{year}{2022}\natexlab{}.
\newblock \showarticletitle{Automatic generation of programming exercises and
  code explanations using large language models}. In
  \bibinfo{booktitle}{\emph{Proceedings of the 2022 ACM Conference on
  International Computing Education Research-Volume 1}}.
  \bibinfo{pages}{27--43}.
\newblock


\bibitem[Singh et~al\mbox{.}(2013)]%
        {singh2013automated}
\bibfield{author}{\bibinfo{person}{Rishabh Singh}, \bibinfo{person}{Sumit
  Gulwani}, {and} \bibinfo{person}{Armando Solar-Lezama}.}
  \bibinfo{year}{2013}\natexlab{}.
\newblock \showarticletitle{Automated feedback generation for introductory
  programming assignments}. In \bibinfo{booktitle}{\emph{Proceedings of the
  34th ACM SIGPLAN conference on Programming language design and
  implementation}}. \bibinfo{pages}{15--26}.
\newblock


\bibitem[Smetsers-Weeda and Smetsers(2017)]%
        {smetsers2017problem}
\bibfield{author}{\bibinfo{person}{Renske Smetsers-Weeda} {and}
  \bibinfo{person}{Sjaak Smetsers}.} \bibinfo{year}{2017}\natexlab{}.
\newblock \showarticletitle{Problem solving and algorithmic development with
  flowcharts}. In \bibinfo{booktitle}{\emph{Proceedings of the 12th Workshop on
  Primary and Secondary Computing Education}}. \bibinfo{pages}{25--34}.
\newblock


\bibitem[Sykes(2010)]%
        {sykes2010design}
\bibfield{author}{\bibinfo{person}{Edward~R Sykes}.}
  \bibinfo{year}{2010}\natexlab{}.
\newblock \showarticletitle{Design, Development and Evaluation of the Java
  Intelligent Tutoring System.}
\newblock \bibinfo{journal}{\emph{Technology, Instruction, Cognition \&
  Learning}} \bibinfo{volume}{8}, \bibinfo{number}{1} (\bibinfo{year}{2010}).
\newblock


\bibitem[Tang et~al\mbox{.}(2023)]%
        {tang2023empirical}
\bibfield{author}{\bibinfo{person}{Ningzhi Tang}, \bibinfo{person}{Meng Chen},
  \bibinfo{person}{Zheng Ning}, \bibinfo{person}{Aakash Bansal},
  \bibinfo{person}{Yu Huang}, \bibinfo{person}{Collin McMillan}, {and}
  \bibinfo{person}{Toby Jia-Jun Li}.} \bibinfo{year}{2023}\natexlab{}.
\newblock \showarticletitle{An Empirical Study of Developer Behaviors for
  Validating and Repairing AI-Generated Code}. Plateau Workshop.
\newblock


\bibitem[Vaithilingam et~al\mbox{.}(2022)]%
        {vaithilingam2022expectation}
\bibfield{author}{\bibinfo{person}{Priyan Vaithilingam},
  \bibinfo{person}{Tianyi Zhang}, {and} \bibinfo{person}{Elena~L Glassman}.}
  \bibinfo{year}{2022}\natexlab{}.
\newblock \showarticletitle{Expectation vs. experience: Evaluating the
  usability of code generation tools powered by large language models}. In
  \bibinfo{booktitle}{\emph{Chi conference on human factors in computing
  systems extended abstracts}}. \bibinfo{pages}{1--7}.
\newblock


\bibitem[Wang et~al\mbox{.}(2022)]%
        {wang2022exploring}
\bibfield{author}{\bibinfo{person}{Wengran Wang}, \bibinfo{person}{Audrey
  Le~Meur}, \bibinfo{person}{Mahesh Bobbadi}, \bibinfo{person}{Bita Akram},
  \bibinfo{person}{Tiffany Barnes}, \bibinfo{person}{Chris Martens}, {and}
  \bibinfo{person}{Thomas Price}.} \bibinfo{year}{2022}\natexlab{}.
\newblock \showarticletitle{Exploring Design Choices to Support Novices'
  Example Use During Creative Open-Ended Programming}. In
  \bibinfo{booktitle}{\emph{Proceedings of the 53rd ACM Technical Symposium on
  Computer Science Education V. 1}}. \bibinfo{pages}{619--625}.
\newblock


\bibitem[Xu et~al\mbox{.}(2022)]%
        {xu2022ide}
\bibfield{author}{\bibinfo{person}{Frank~F Xu}, \bibinfo{person}{Bogdan
  Vasilescu}, {and} \bibinfo{person}{Graham Neubig}.}
  \bibinfo{year}{2022}\natexlab{}.
\newblock \showarticletitle{In-ide code generation from natural language:
  Promise and challenges}.
\newblock \bibinfo{journal}{\emph{ACM Transactions on Software Engineering and
  Methodology (TOSEM)}} \bibinfo{volume}{31}, \bibinfo{number}{2}
  (\bibinfo{year}{2022}), \bibinfo{pages}{1--47}.
\newblock


\bibitem[Xu et~al\mbox{.}(2019)]%
        {xu2019block}
\bibfield{author}{\bibinfo{person}{Zhen Xu}, \bibinfo{person}{Albert~D
  Ritzhaupt}, \bibinfo{person}{Fengchun Tian}, {and}
  \bibinfo{person}{Karthikeyan Umapathy}.} \bibinfo{year}{2019}\natexlab{}.
\newblock \showarticletitle{Block-based versus text-based programming
  environments on novice student learning outcomes: A meta-analysis study}.
\newblock \bibinfo{journal}{\emph{Computer Science Education}}
  \bibinfo{volume}{29}, \bibinfo{number}{2-3} (\bibinfo{year}{2019}),
  \bibinfo{pages}{177--204}.
\newblock


\end{thebibliography}
